\documentclass[twocolumn, prl]{revtex4}

\usepackage{graphicx}
\usepackage{dcolumn}
\usepackage{bm}

\begin{document}

\title{ Critical Exponents  in a Quantum Phase Transition of an Anisotropic
2D  Antiferromagnet}

\author {S. Moukouri}

\affiliation{ Department of Physics and  Michigan Center for 
          Theoretical Physics \\
         University of Michigan, 2477 Randall Laboratory, Ann Arbor MI 48109}

\begin{abstract}
I use the two-step density-matrix renormalization group method to
extract the critical exponents $\beta$ and $\nu$ in the transition
 from a N\'eel $Q=(\pi,\pi)$ phase to a magnetically  
disordered phase with a spin gap. I find that the exponent $\beta$ computed
from the magnetic side of the transition is consistent with that
of the classical Heisenberg model, but not the exponent $z\nu$ computed 
from the disordered side.  I also show the contrast 
between integer and half-integer spin cases.
\end{abstract}

\maketitle

There is  current interest in studying quantum phase transitions (QPT) in
frustrated  quantum systems \cite{sachdev1}. An interesting phenomenon 
generated by frustration is the dimensional reduction recently observed
in the Bose-Einstein condensation QPT of $BaCuSi_2O_6$ \cite{batista}. 
In addition
to their connection to real materials, these 2D models could be an example 
 of systems where the conventional Landau-Ginzburg-Wilson (LGW) approach
to phase transitions may fail. Of particular interest is the transition
from an antiferromagnet (AFM) to a disordered phase with a spin gap and 
short-range magnetic correlations known as a valence bond crystal (VBC). 
A very fruitful approach to the study of these transitions relies on to the 
mapping of the Heisenberg model to the non-linear sigma (NL$\sigma$) 
model\cite{sachdev2,chakravarty}. If the Berry phase terms are neglected 
at the transition as suggested in Ref.\cite{sachdev2}, the critical 
behavior of the quantum 2D model is identical to that of the NL$\sigma$ model. 
However, the Berry phase effects are absent only in the AFM phase; they are 
important in the disordered phase and they might well modify the critical 
exponents of the 2D Heisenberg model as suggested in Ref.\cite{chakravarty}.

 The transition from an AFM to a spin gap in the 2D Heisenberg model 
was studied in Ref.\cite{troyer} by the Quantum Monte Carlo (QMC) method. 
It was concluded that this transition belongs to the universality class
of the $O(3)$ classical Heisenberg model as may be expected from the
$NL\sigma$ mapping. In this study, the disordered phase was generated
by the explicit dimerization of the bonds. This would support the view 
of Ref.\cite{sachdev2} that Berry phase effects do not affect the
transition. But due to the minus sign problem, the important case of
transition driven by frustration cannot be studied by the QMC method. 
 Hence, it remains to be seen whether frustration merely brings some 
technical complication, as would suggest the hypothesis that the universality
class does not depend on the detail of the Hamiltonian, or it can drive the transition to another universality class.

There is also a more general interest in studying frustration induced
transitions. The conventional approach to quantum phase transitions is to 
use the path integral formalism  which maps quantum phase transitions in 
dimension $d$ to classical transitions in dimension $d+1$ to which the 
machinery of the Landau-Ginzburg-Wilson theory is then applied. However, 
this mapping is fraught with difficulty in some important situations. 
For some experimentally relevant quantum mechanical models, the corresponding 
classical functional integrals have non-positive or even  
complex valued Boltzmann weights.  This would suggest in such cases that 
the quantum transition does not have a classical equivalent. It is thus 
important to study the quantum models directly. But the direct study of such 
quantum mechanical models has proven to be a
formidable challenge to condensed-matter theorists both numerically and
analytically. In particular, the QMC method, which in its 
formulation uses this quantum-classical mapping, is at present 
unable to address these issues because of the non-positive values of the
effective Boltzmann weights. This state of matter has stalled progress
in the understanding of QPT.

 In a recent publication \cite{moukouri-TSDMRG2}, 
it  has been shown that 2D problems may be attacked by a chain perturbation
theory method. This method called the two-step DMRG has recently been 
applied to study quantum phase transitions induced by frustration in two 
dimensions. The two-step DMRG  is not a simple Green's function 
perturbation expansion which is known to fail in situations where quantum
fluctuations are important. It is rather a perturbation expansion on the 
reduced Hamiltonian. It thus retains the full low-energy many-body dynamics 
of the original problem. The two-step method has the ability to  
reach an ordered phase when it exists. In this letter, we report the 
computation of the critical exponents in  the important case of the 
transition from an AFM to a VBC induced by frustration for $S=1$. Such a 
transition was suggested in Ref.\cite{sachdev3} from a large $N$ 
analysis and  found in an anisotropic 2D Heisenberg model with $S=1$ in 
Ref.\cite{moukouri-TSDMRG2}. In addition we highlight the difference 
between integer spin and half-integer spin systems by comparing the
case of $S=\frac{3}{2}$ to that of $S=1$.

 I study the following Heisenberg model on the anisotropic square
lattice:  

\begin{eqnarray}
 \nonumber H=J_{\parallel} \sum_{i,l}{\bf S}_{i,l}{\bf S}_{i+1,l}+J_{\perp} \sum_{i,l}{\bf S}_{i,l}{\bf S}_{i,l+1}\\
+J_d \sum_{i,l}({\bf S}_{i,l}{\bf S}_{i+1,l+1}+{\bf S}_{i+1,l}{\bf S}_{i,l+1}),
\label{hamiltonian}
\end{eqnarray}

\noindent where $S=1$, $J_{\parallel}$ is the in-chain exchange parameter 
and is set to 1; $J_{\perp}$ and $J_d$ are respectively the transverse and 
diagonal interchain exchanges. 

In the TSDMRG, we start by applying the DMRG 
to a single chain for which we obtain the low energy eigenvalues 
$\epsilon_n$ and eigenvectors $|\phi_n\rangle$. Then the 
Hamiltonian (\ref{hamiltonian}) is projected unto the tensor 
product $|\Phi_{\parallel [n]}\rangle =  |\phi_{n_1}\rangle  |\phi_{n_2}\rangle ...|\phi_{n_L} \rangle$ yielding the effective low
energy Hamiltonian

\begin{eqnarray}
 \nonumber \tilde{H} \approx \sum_{[n]} E_{\parallel [n]} |\Phi_{\parallel [n]}
\rangle \langle\Phi_{\parallel [n]}| +
\nonumber J_{ \perp} \sum_{il} {\bf \tilde{S}}_{i,l} {\bf \tilde{S}}_{i,l+1}
 +\\
J_{d} \sum_{il} {\bf \tilde{S}}_{i,l} {\bf \tilde{S}}_{i+1,l+1}+
{\bf \tilde{S}}_{i+1,l} {\bf \tilde{S}}_{i,l+1},
\label{effh}
\end{eqnarray}

\noindent where  $E_{\parallel [n]}$ is the sum of eigenvalues of the
different chains, $E_{\parallel[n]}=\sum_l{\epsilon_{n_l}}$;
 ${\bf \tilde{S}}_{i,l}$ are the  renormalized 
matrix elements in the single chain basis. They are given by
${\bf \tilde{S}}_{i,l}^{n_l,m_l}=\langle \phi_{n_l}|
{\bf S}_{i,l}|\phi_{m_l}\rangle$. Since the diagonalization has been 
made in the direction of the chains, the effective Hamiltonian(\ref{effh}) is
1D. I again use the DMRG to obtain its spectrum. 

The TSDMRG has been extensively checked\cite{moukouri-TSDMRG2,alvarez-TSDMRG}, 
and is variational. It is controlled by two parameters $m_1$ and $m_2$ which
are the numbers of states kept during the first and second step respectively.
$m_1$ should be large enough so that not only the ground-state energy of a
single chain $l$ but all the $\epsilon_{n_l}$ and $\phi_{n_l}$ retained to
be used during the second step are accurate enough. And, $m_2$ should be 
large enough so that $J_{\perp},~J_d \ll \epsilon_{n_l}-\epsilon_{0_l}$. 
For instance for $S=\frac{1}{2}$, we find that the ground-state energy 
per site of a $16 \times 17$ system evolves from $-0.43481$ when 
$(m_1,m_2)= (128,80)$ to $-0.43681$ when $(m_1,m_2)=(256,96)$ these energies
are to be compared to the QMC energy $-0.43529$. It may be argued that 
despite this accuracy in the ground-state energy, the TSDMRG would not
couple the chains effectively and thus may essentially retain 1D physics.
This would result in a rapid decay of the tranverse spin-spin correlation
function $C_{\perp}$. To counter this argument we show in Table\ref{vsqmc} 
 $C_{\perp}$ for the $16 \times 17$ systems. When $m_1$ and $m_2$ are
large enough, there is no spurious decay of the correlation of the TSDMRG,
they appear to decay even slower than the QMC ones. Hence, although it
starts from an isolated chain, the TSDMRG is able to describe the 2D
regime.

In Ref(\cite{moukouri-TSDMRG3}), it was found that as $J_d$ moves towards 
the maximally frustrated point $J_d \approx J_{\perp}/2$ the
2D system progressively relax to nearly disconnected chains. At this point,
the ground state energy is nearly equal to that of disconnected chains, 
the transverse spin-spin correlations decay exponentially, and the transverse
bond-strength is equal to zero up to the numerical accuracy. Hence, for 
integer spins, the nature of the ground state is determined by the competition 
between $J_{eff}=J_{\perp}-2J_d$ which favors a magnetic phase and the 1D 
Haldane gap $\Delta_H$ which favors a disordered phase.  For half-integer
spin systems, the system would be ordered everywhere except exactly at
the maximally frustrated point where it would be critical.

\begin{table}
\begin{ruledtabular}
\begin{tabular}{cccc}
 $l$ & $(m_1,m_2)=(128,80)$ & $(m_1,m_2)=(256,96)$ & QMC  \\
\hline
 1 & -0.02116& -0.03022 & -0.02533(1) \\
 2 & 0.00726 & 0.01088 &  0.00854(1)\\
 3 & -0.00320 & -0.00572 & -0.00399\\
 4 & 0.00147 & 0.00320 & 0.00201\\
 5 & -0.00078 & -0.00186 & -0.00105\\
 6 & 0.00030 & 0.00108 & 0.00056\\
 7 & -0.00013 & -0.00063 & -0.00030\\
 8 & 0.00006 & 0.00036 &  0.00015\\
\end{tabular}
\end{ruledtabular}
\caption{ Transverse correlation $C_{\perp}(l)$ for a $16 \times 17$
system with $J_{\perp}=1$ and $J_d=0$  of two sets of
TSDMRG parameters against QMC. The origin is taken at the middle 
of the lattice.}
\label{vsqmc}
\end{table}

This behavior predicted by TSDMRG may be found analytically by 
applying the Haldane \cite{haldane, arovas} mapping of the 
Heisenberg model to the non-linear $\sigma$-model. I write

\begin{eqnarray}
{\hat{\bf {\Omega}}}_{jl}=\eta_j {\hat{\bf n}}_{jl}\sqrt{1-{\bf m}_{jl}^2}+
{\bf m}_{jl},
\end{eqnarray}

\noindent where ${\hat{\Omega}}_{jl}.{\bf S}_jl|{\hat{\Omega}}_{jl}\rangle =S
|{\hat{\Omega}}_{jl}\rangle$ is a spin-coherent state, ${\hat{n}}_{jl}$ is the
local N\'eel field, $\eta_i$ is the sublattice modulation, 
${\bf m}=\frac{v_0 {\bf l}}{S\hbar}$ and $\bf l$ describes the ferromagnetic 
fluctuations about the N\'eel order and $v_0$ is the unit cell volume. Keeping
only up to second order terms in ${\bf m}_{jl}$, the 
Hamiltonian(\ref{hamiltonian}) becomes,

\begin{eqnarray}
\nonumber H'=\sum_l \int dx [\rho_s(\partial_x {\hat {\bf n}}_l(x))^2+ 
          \chi^{-1}{\bf m}_l^2(x)-(J_{\perp}-2J_d)\times\\
{\hat {\bf n}}_l(x).{\hat {\bf n}}_{l+1}(x)+
         (J_{\perp}+2J_d){\bf m}_l(x).{\bf m}_{l+1}(x)],
\end{eqnarray}
       
\noindent where $\rho_s=-\frac{S^2}{2Nv_0}\sum_{jj'}\eta_j\eta_j'(x_j-x_{j'})^2$ and
$\chi^{-1}=\frac{v_0}{2N\hbar^2}\sum_{jj'}(1-\eta_j\eta_{j'})$ are 
respectively the spin stiffness and the inverse susceptibility along 
the chains. As usual, one must add the Berry phase term,

\begin{eqnarray}
S_B=S\sum_{jl}\eta_j\omega[{\hat {\bf n}}_{jl}]+\sum_l\int dx 
{\bf m}_l(x).\frac{\partial{\hat {\bf n}}_l(x)}{\partial t} \times 
{\hat {\bf n}}_l(x).
\end{eqnarray}
 
Then, writing the total action resulting from $H'$ and $S_B$ and performing 
the integration over the fields ${\bf m}_l(x)$ with the constraint 
${\hat {\bf n}}_l(x).{\bf m}_l(x)=0$
yields, at the maximally frustrated point, the following effective action:

\begin{eqnarray}
S_E=\sum_l \int dx \rho_s(\partial_x {\hat {\bf n}}_l^2(x)+K_{ll'}
(\frac{\partial {\hat {\bf n}}_l}{\partial t}\times {\hat {\bf n}}_l).
(\frac{\partial {\hat {\bf n}}_{l'}}{\partial t}\times {\hat {\bf n}}_{l'}),
\end{eqnarray}

\noindent where $K^{-1}_{ll'}=\delta_{ll'}\chi^{-1}+(J_{\perp}+2J_d)\delta_{ll'+1}$. 
Hence, to the leading order, at $J_d=J_{\perp}/2$, the chains are only coupled
by terms originating from the Berry phase on each chain. Therefore as found
numerically \cite{moukouri-TSDMRG2}, unlike the magnetic case where $S$ does
not play any role, the physics of the 2D system near the
maximally frustrated point will strongly depends on the 1D physics, i.e.,
whether $S$ is integer or half-integer as found numerically.

\begin{figure}
\begin{center}
$\begin{array}{c@{\hspace{0.5in}}c}
         \multicolumn{1}{l} {}\\ [-0.23cm]
\includegraphics[width=4.65cm, height=5.25cm]{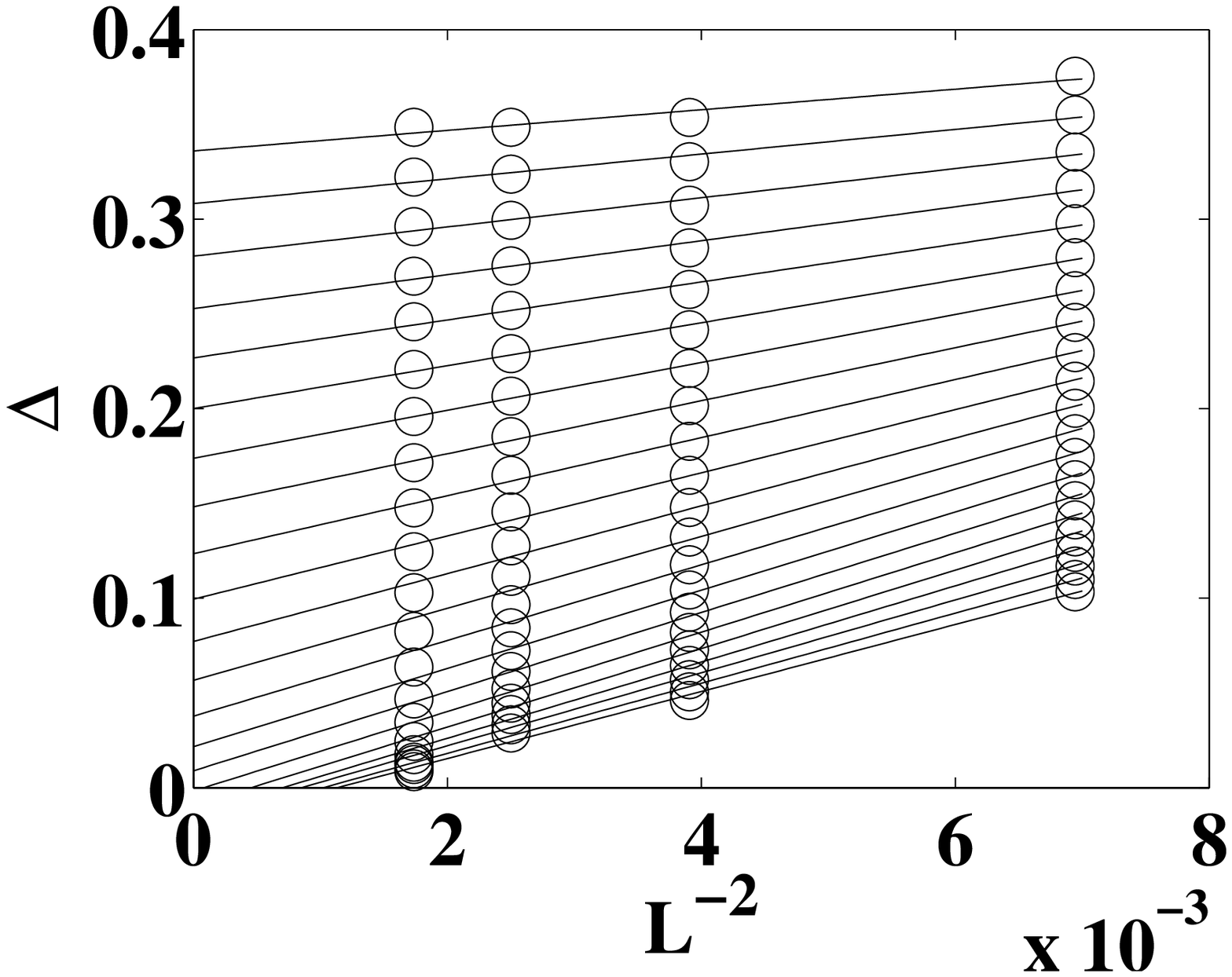}
\hspace{0.005cm}
\vspace{0.6cm}
\includegraphics[width=4.65cm, height=5.25cm]{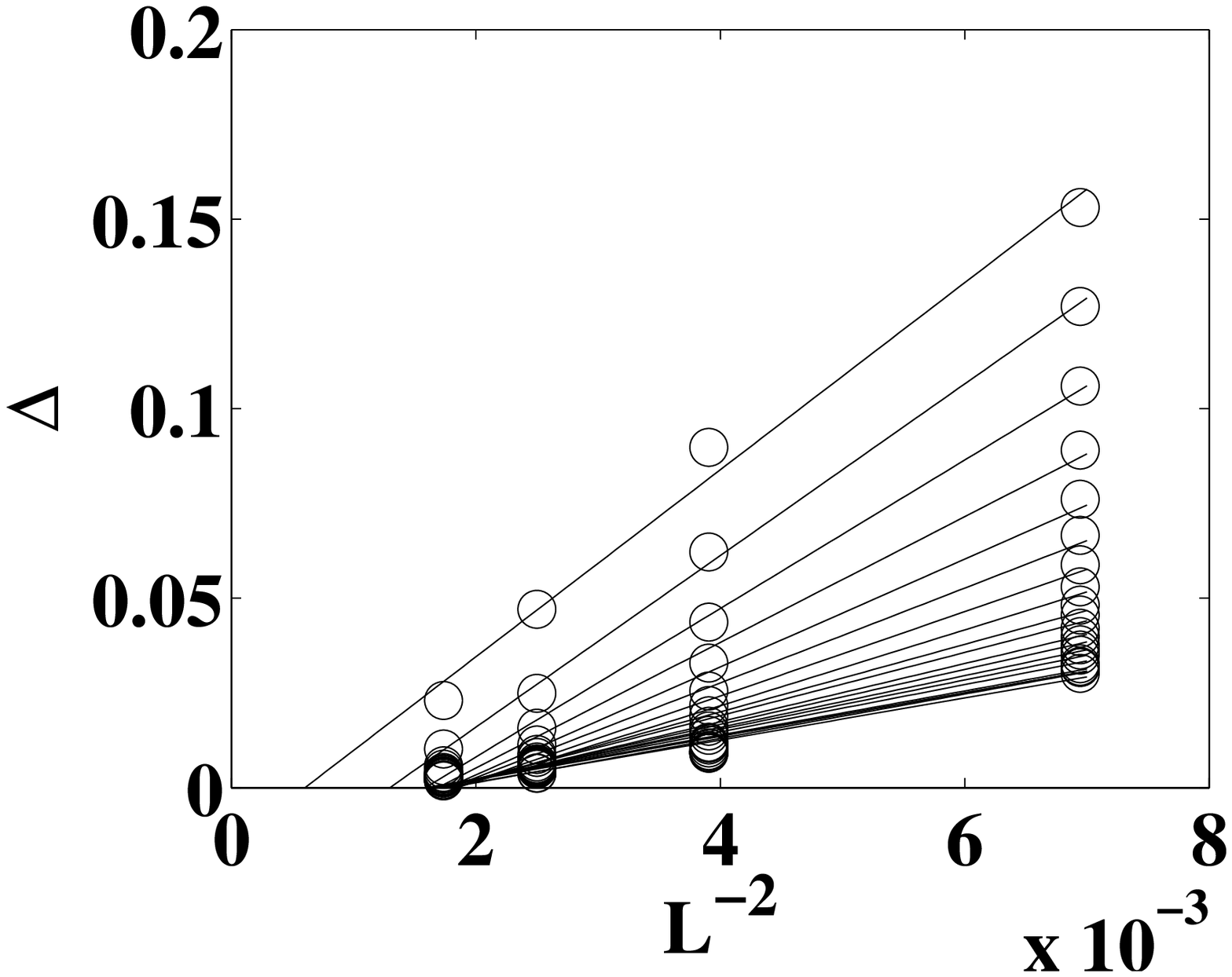}
\end{array}$
\includegraphics[width=3. in, height=2. in]{gap2.eps}
\end{center}
\caption{TOP: Gaps for $S=1$ (left) and $S=\frac{3}{2}$ (right) for 
$J_{\perp}=0.2$ and for $J_d$ ranging from
$J_d=0.06$ (bottom) to $J_d=0.1$ (top) with the step $\delta J_d=0.02$.
 BOTTOM: Extrapolated gap for $S=1$ and for the same values of $J_d$ as above.}
\label{gapll}
\end{figure}

 For relatively small $J_{\perp}$, the eventual magnetic order parameter is
 roughly $m \propto \sqrt{J_{\perp}}$ for $J_d=0$ \cite{shulz} and $m$ 
will get even smaller when $J_d \neq 0$. For this reason, the critical 
behavior of $S=\frac{1}{2}$ systems is very difficult to study. I therefore  
choose $S=1$ and $S=\frac{3}{2}$ for which $m$ is larger and could be 
extrapolated to from lattice sizes not too large. The simulations were 
done on  $L\times(L+1)$ systems with $L=8$, $12$, $16$, $20$ and $24$. 
I use the periodic boundary
conditions (PBC) in the direction parallel to the chains and open boundary
conditions (OBC) in the transverse direction. I kept a maximum of 
$m_1=243$ states during the first step. 
I targeted the spin sectors $S^z=0,\pm1,\pm2$. The maximum truncation error 
during this step was $\rho_1=9 \times 10^{-6}$. During the second step I 
kept a maximum of $m_2=100$ states.  During the second step I targeted 
two states $S^z=0,1$. The truncation error was about $5\times10^{-4}$ in 
the magnetic phase. It dropped to $6\times10^{-7}$ in the disordered phase. 
The TSDMRG is thus at its best in the vicinity of the transition and in the 
disordered phase. 

In Fig.\ref{gapll}, I show the finite size gap $\Delta$ 
for $S=1$ and $S=\frac{3}{2}$.
In all cases, $J_{\perp}=0.2$ and $J_d$ is varied from $0.06$ to $0.1$.
The $S=1$ and $S=\frac{3}{2}$ show a striking difference. For $S=1$, starting
from $0.06$ where the system is gapless and decay to $0$ faster than $L^{-1}$,
a gap opens around $J_d=0.073$. This gap opening signals a transition from an 
AFM to a magnetically disordered phase. On the other hand, the $S=\frac{3}{2}$ 
system remains gapless for all values of $J_d$. The fact that $S=\frac{3}{2}$
remains gapless makes the transition at the maximally frustrated point
quite difficult to study because long-range order vanishes only at the
critical point an beyond this point, the N\'eel phase  $(\pi,0)$ sets in. 
This is complicated by the fact that for finite systems with OBC, the
maximally frustrated point is not exactly at $J_d=J_{\perp}/2$ 
\cite{moukouri-TSDMRG2}. Hence, the critical behavior could only be analyzed
for $S=1$. I show the in Fig.\ref{gapll} extrapolated gap as function of 
$J_d$ for $S=1$. $\Delta$ vanishes at $J_{d_c} \approx 0.073$. 
Taking as granted the large $N$ prediction that the transition is of
second order, I extracted  the critical exponent 
$z\nu=1.205$ ($\Delta \propto (J_d-J_{d_c})^{z\nu}$).  
 Hence, Hamiltonian (\ref{hamiltonian}) does not belong to the universality 
class of the classical $O(3)$ Heisenberg model for which  
$z\nu=0.7048\pm0.0030$.

\begin{figure}
\begin{center}
$\begin{array}{c@{\hspace{0.5in}}c}
         \multicolumn{1}{l} {}\\ [-0.23cm]
\includegraphics[width=4.65 cm, height=5.25cm]{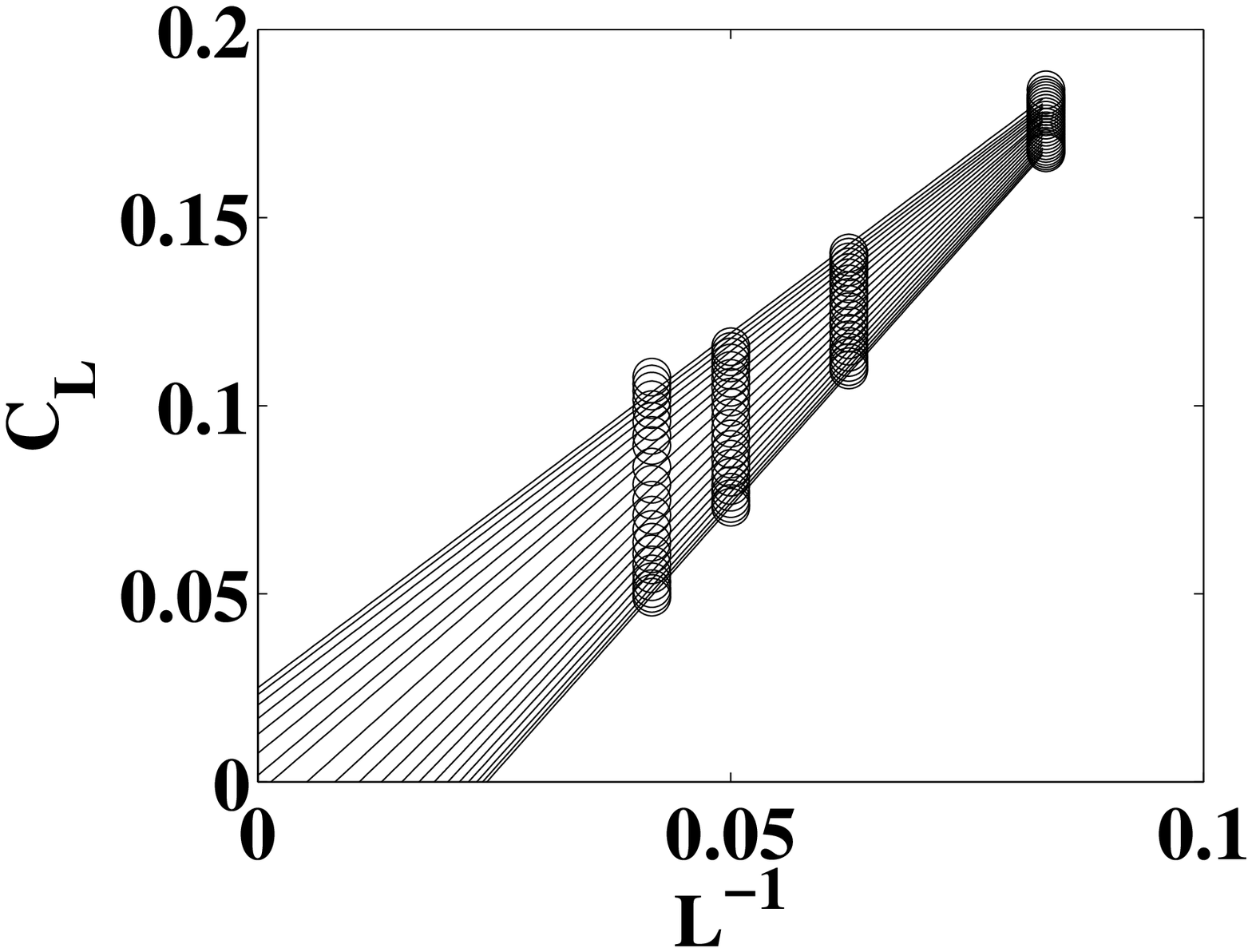}
\hspace{0.005cm}
\vspace{0.5cm}
\includegraphics[width=4.65 cm, height=5.25cm]{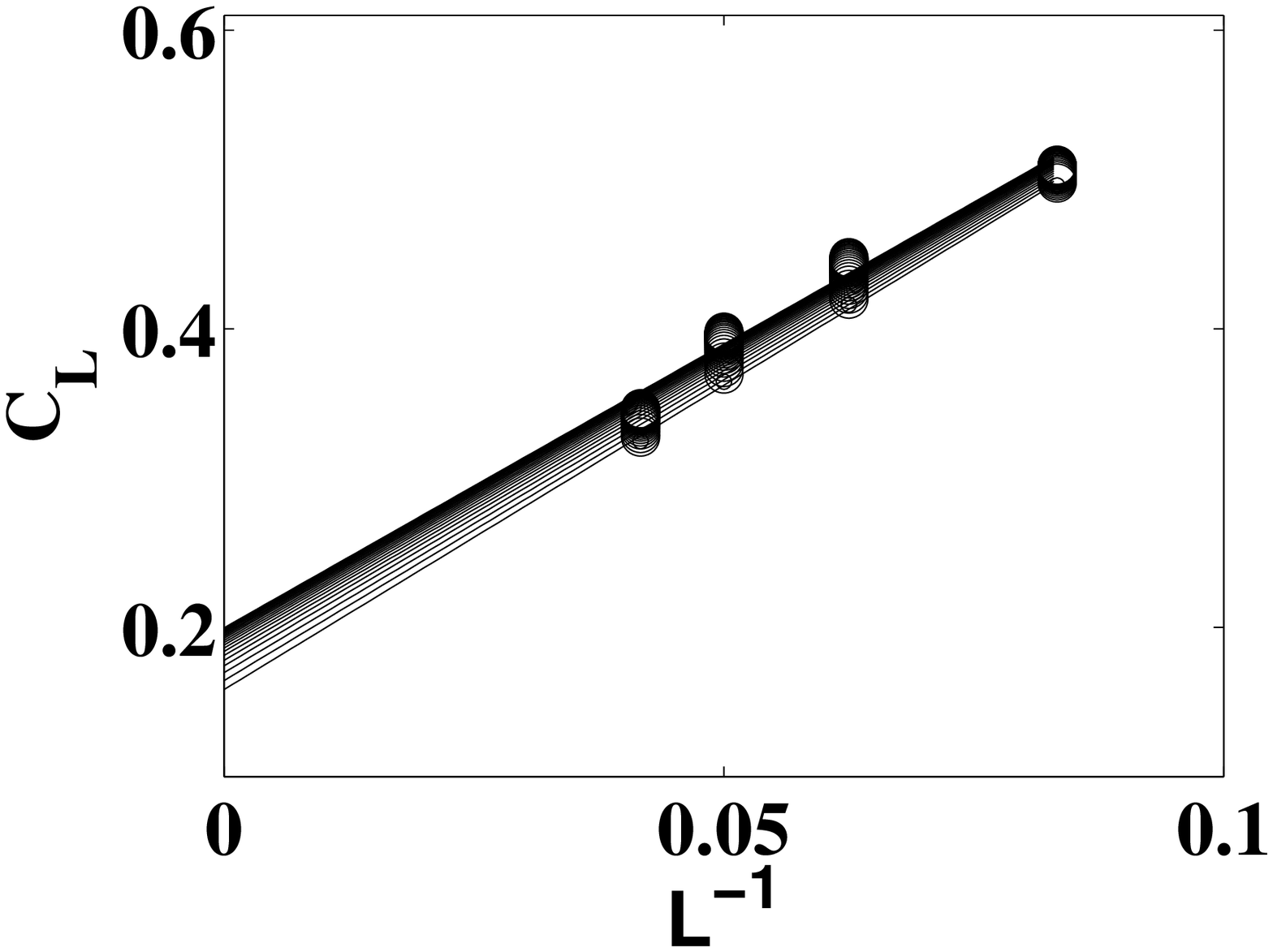}
\end{array}$
\end{center}
\includegraphics[width=3. in, height=2. in]{orderpar.eps}
\caption{ TOP: End-to-center correlation $C_L$ for $S=1$ (left) and 
$S=\frac{3}{2}$ (right) for $J_{\perp}=0.2$ and $J_d$ ranging from
$J_d=0.06$ (bottom) to $J_d=0.1$ (top) with the step $\delta J_d=0.02$.
BOTTOM: Extrapolated magnetization for $S=1$ and for the same values of 
$J_d$ as above.}
\label{orderpar}
\end{figure}

The center-to-end spin-spin correlation function 
$C_L=\frac{1}{3} \langle {\bf S}_{L/2,L/2+1}{\bf S}_{L,L/2+1}\rangle$
is shown in Fig.\ref{orderpar}.  $C_L$ is also dependent on $S$.
For $S=1$, $C_L$ first decays slower than $L^{-1}$ and extrapolates to
a finite value for small $J_d$, then decays
faster than $L^{-1}$ for larger $J_d$, where it extrapolates to $0$. 
For $S=\frac{3}{2}$ for all values of the coupling studied, $C_L$ extrapolates
to a finite value.
The behavior of the magnetization $m=\sqrt{3C_\infty}$ shown in 
Fig.\ref{orderpar} is consistent with that of $\Delta$. For $S=1$, starting
from the AFM phase for $J_d=0.06$, $m$ 
vanishes around $J_d \approx 0.076$. The best fit to data yields the exponent 
$m\propto (J_{d_c}-J_d)^\beta$, $\beta=0.3653$ which is in good 
agreement with  that of the classical Heisenberg model 
$\beta=0.3639 \pm 0.0035$. By contrast, $m$ for $S=\frac{3}{2}$ extrapolates
to a finite value for all values of $J_d \alt 0.1$. This result, with that
seen for $\Delta$, contradicts the large $N$ prediction that there is
a $VBC$ phase in this regime for $S=\frac{3}{2}$ as well. Ultimately, I 
find that long-range order vanishes when $J_d \approx 0.11$ for all 
$L$ studied and, immediately, the systems jumps to the N\'eel phase 
with $Q=(\pi,0)$. The critical behavior for 
$S=\frac{3}{2}$ was quite difficult to study. This is because, close
enough to the critical point, for a fixed $J_d$, starting from the
N\'eel phase with $Q=(\pi,\pi)$, the system evolves to the $Q=(\pi,0)$
N\'eel phase at larger $L$. For this reason, the extrapolations cannot 
 be reliably made.

A previous Monte Carlo simulation \cite{troyer} for a Heisenberg model 
with $S=1/2$ on the $CaV_4O_9$ lattice with found $\nu=0.685 $ and  
$\beta=0.345$. These exponents are in agreement with the classical 
Heisenberg model within statistical errors.   It is not definitive whether 
the QMC results are in contradiction with my results. In the QMC study, 
following the assumption that critical exponents are the same in either 
side of the transition, $\nu$ was calculated from the ordered side through 
the spin stiffness $\rho_s \propto (g_c-g)^{z\nu}$. 
One of the standard results of the classical critical phenomena is that
the critical exponents of the region above the transition are identical
to those below the transition.  It is not however obvious
that this result extends to quantum phase transitions.
From the QMC results and mine, it seems that the exponents found
for model (\ref{hamiltonian}) violate this LGW classical behavior.
 The exponents  in the AFM phase verify the relation
$(d+z-2+\eta)\nu=2\beta$. Since $d=2$, if I assume that the critical exponent
on either side of the transition are identical, I may use $z\nu$ found in the
disordered case in this relation. Then, it is clear the above relation 
between critical exponents is not satisfied given  that $\eta$
is predicted to be small, $\eta=0.0033$ for the classical Heisenberg model.
This violation is possibly related to a dimensional reduction at the
critical point. At the maximally frustrated point the bond strength in the
transverse direction vanishes. Hence, the system is effectively 1D. It could
be that this 1D physics emerges at the critical point.
An alternative possibility which is consistent with the small difference
find in $J_{d_c}$ from $\Delta$ and from $m$, is that the transition is of
first order and there is a narrow region of coexistence of the
two phases.  

\begin{acknowledgments}
I am very grateful to Dan Arovas for very helpful exchanges. 
This work was supported by the NSF Grant No. DMR-0426775.
\end{acknowledgments}


\begin{thebibliography}{99}
\bibitem{sachdev1}  S. Sachdev "Quantum Phase Transitions", Cambridge University
 Press  (1999).
\bibitem{batista} S.E. Sebastian, N. Harrison, C.D. Batista, L. Balicas,
                 M. Jaime, P.A. Sharma, N. Kawashima and I.R. Fisher,
                 Nature {\bf 441}, 617 (2006).
\bibitem{troyer} M. Troyer, M. Imada, and K. Ueda, J. Phys. Soc. Jpn {\bf 66},
          2957 (1997).
\bibitem{moukouri-TSDMRG2} S. Moukouri, Phys. Rev. {\bf B 70}, 014403 (2004).
\bibitem{moukouri-TSDMRG3} S. Moukouri,  J. Stat. Mech. P02002 (2006).
\bibitem{alvarez-TSDMRG} J.V. Alvarez and S. Moukouri, Int. J. Mod. Phys.
                         {\bf C 16}, 84 (2004).
\bibitem{haldane}F.D.M. Haldane, Phys. Rev. Lett. {\bf 61}, 1029 (1988).
\bibitem{arovas} D. Arovas, Boulder Summer School lectures notes (2003,
              unpublished).
\bibitem{sachdev2} A.V. Chubukov, S. Sachdev and J. Ye, Phys. Rev. {\bf B 49},
                   11919 (1994).
\bibitem{chakravarty} S. Chakravarty, B.I. Halperin and D.R. Nelson, 
                     Phys. Rev. {\bf B 39}, 2344 (1989).
\bibitem{shulz} H.J. Schulz, Phys. Rev. Lett. {\bf 77}, 2790 (1996).
\bibitem{sachdev3} N. Read and S. Sachdev, Phys. Rev. Lett. {\bf 66},
                   1773 (1991). 

\end{thebibliography}
\end{document}